\documentclass[twocolumn,aps,amsmath,amssymb,showpacs,floatfix]{revtex4}
\usepackage{graphicx}
\usepackage{dcolumn}
\usepackage{bm}
\usepackage{hyperref} 
\usepackage{latexsym}
\usepackage{float}

\def\av#1{\langle#1\rangle}
\def\tree#1{SF$\perp_{#1}$}
\def\loops#1{SF$\Delta_{#1}$}
\def\Tp{{T'\!\!}}
\def\dw{d_{\text{w}}}
\def\df{d_{\text{f}}}

\def\l{\lambda}

\begin{document}
\title{What is Special about Diffusion on Scale-Free Nets?}   
\author{Erik M. Bollt}
\email{bolltem@clarkson.edu}
\affiliation{Department of Math and Computer Science, Clarkson University, Potsdam, NY
13699-5805}
\affiliation{Department of Physics, Clarkson University,
Potsdam NY 13699-5820}
\author{Daniel ben-Avraham}
\email{benavraham@clarkson.edu}
\affiliation{Department of Physics, Clarkson University,
Potsdam NY 13699-5820}

\begin{abstract} 
  We study diffusion (random walks) on recursive scale-free graphs, and contrast
  the results to similar studies in other analytically soluble media.  This allows us to identify 
  ways in which diffusion in scale-free graphs is special.  Most notably, scale-free architecture
  results in a faster transit time between existing nodes, when the network grows in size;
  and walks emanating from the most connected nodes are recurrent, despite the network's
  infinite dimension.   We also find that other
  attributes of the graph, besides its scale-free distribution, have a strong influence on the
  nature of diffusion.
\end{abstract}

\maketitle

\date{\today}


\section{Introduction}

Recently there has been growing interest in large stochastic networks, such as social networks
of contacts, networks of collaboration between scientists, the Internet, flight connections, networks
of predator-prey, and many other examples in biology, sociology, economics, even
 linguistics~\cite{barabasi,albert02,dorogovtsev02,bornholdt}.  
While the inquiry into networks begins already with 
E\"uler's seminal work on Graph Theory, the latest interest stems from the discovery that many
of the large networks encountered in everyday life seem to be \emph{scale-free}~\cite{barabasi,albert02}:
 the distribution
of the degree of the nodes (the number of links $k$ coming out of a node) has a power-law
tail,
\begin{equation}
P(k)\sim k^{-\lambda},\qquad k\gg1.
\end{equation}
The scale-free degree distribution is implicated in exotic properties of the networks, such
as increased resilience to random failure (the nets maintain a connected backbone under
random dilution of the nodes, when the degree exponent is $2\leq\lambda\leq 3$) and 
increased vulnerability to removal of the most highly connected nodes~\cite{cohen}.  Indeed, it has been
suggested that the need to attain such attributes might have acted, in some cases, as the
natural selection mechanism giving rise to the scale-free distribution in the first place.

In this communication we examine whether there is anything special about transport by diffusion
in scale-free nets.  Diffusion is of course a very natural mode of transport, where hopping from one 
node to the next is unaffected by the history of the walk.  (We think mainly of simple random walks, where
the walker takes one time step to transit between adjacent nodes, and refer to these as diffusion.)
Accordingly, diffusion on various networks has been studied extensively~\cite{perc,fpp,L,dba}
We approach the question by comparing recursive scale-free nets~\cite{dor_rec,rec} to other nets.  
To make the
comparison meaningful, we require that the nets possess a well-defined average
degree, $\av{k}<\infty$,  in the thermodynamic limit of $N\to\infty$~\cite{remark1}.
The \emph{recursive} scale-free nets are studied for convenience, because they allow for 
exact analysis~\cite{kr}.
Our hope is that stochastic scale-free nets are well represented by the examples chosen, though
to which extent they do remains an open question.
We do find advantages to a scale-free architecture in the form of 
faster transit times between existing nodes, when the network grows in size.  We also observe that
walks emanating from the most connected nodes are recurrent, despite the networks' infinite dimension.

The rest of this paper is organized as follows.
Diffusion and its effectiveness can be gauged in many different ways, and in Section~\ref{measures}
we explain which aspects of diffusion we choose to focus on.  As an example we examine
the complete graph, thereby introducing some basic techniques used later in the study of scale-free
nets.  In Section~\ref{other} we briefly survey how diffusion fares on other network architectures,
including regular and fractal lattices.  The main bulk of our results is given in Section~\ref{results},
where we analyze two recursive scale-free nets, sharing the same degree distribution, but 
representing two extremes in terms of the number of paths connecting any two nodes.  Our 
conclusions are summarized in Section~\ref{discussion}.  Technical details about the
scale-free networks studied here, and the methods used for this purpose, are left to the appendices.

\section{Measures of diffusion}
\label{measures}

Diffusion along networks can be characterized in many different ways, but we find it most
convenient to focus on the following:
\begin{itemize}

\item Mean transit time, $T_{ij}$:  This is the mean first passage time (FPT) between
two distinct nodes $i$ and $j$, averaged over all node pairs.

\item Mean return time, $T_{ii}$:  This is the mean time for returning to a node $i$
for the first time (after having left it), averaged over all the nodes of the net.

\item Scaling efficiency:  How the various FPTs increase upon growth of the net.

\item Recurrence:  Diffusion from a node is recurrent if a walker starting from the node
returns to the node almost surely over the course of time.  When diffusion
is not recurrent, we call it \emph{transient}.  The concept is useful only
in the limit $N\to\infty$, since all nodes are recurrent when $N<\infty$, regardless of
the net's architecture.
\end{itemize}

The mean transit time $T_{ij}$ is a measure of how fast one can navigate the net randomly,
through diffusion.  The mean return time $T_{ii}$ is a rough measure 
of \emph{prominence}: how easy it is to find your way back home.
The average over all $ij$ pairs of nodes (for $T_{ij}$), or over all nodes $i$ (for $T_{ii}$)
is necessary in order to obtain a single figure of merit representative of the graph as a whole.
We omit the commonly used angular brackets denoting average, for brevity, whenever there 
is no risk of confusion.
In situations where the graph is highly inhomogeneous the average
conceals the wide distribution found when looking at individual nodes.  We shall indeed
examine this distribution for the notoriously inhomogeneous case of scale-free nets.

A quantity related to the mean transit time is the mean sojourn time, the FPT from node $i$
to node $j$ and back: $T_{iji}=T_{ij}+T_{ji}$.  While $T_{ij}$ is not necessarily
equal to $T_{ji}$ for a specific pair of nodes, upon taking the average over all pairs
one simply has $\av{T_{iji}}=2\av{T_{ij}}$, so the mean sojourn time is trivially related
to the mean transit time.  It will be ignored in our study.  Note  that there is no simple relation
between $T_{iji}$ and $T_{ii}$, and separate consideration of the latter is necessary.

In all cases, the relevant question is how the FPT scales with the size of the net, $N$.  Clearly,
the various FPTs increase with $N$, but we ask whether there is a preferred 
architecture that minimizes the increase.  To tackle this issue, we introduce the concept
of \emph{scaling efficiency}.
Suppose that the net grows in size, from $N$ to $bN$ nodes.  We expect that in the thermodynamic
limit of $N\to\infty$
\begin{equation}
T_{ij}(bN)\sim b^{\eta_{ij}}T_{ij}(N),
\end{equation}
defining the scaling efficiency exponent $\eta_{ij}$.  (An analogous relation for $T_{ii}$
defines $\eta_{ii}$.) 

Scaling efficiency can be restricted to the nodes already present in the network \emph{before}
growth.  Let $\Tp_{ij}(bN)$ be the mean transit time in the augmented network,
averaged over the \emph{original} set
of nodes (before growth).  Then we define the \emph{restricted} scaling efficiency exponent,
$\nu_{ij}$, thus:
\begin{equation}
\Tp_{ij}(bN)\sim b^{\nu_{ij}}T_{ij}(N),
\end{equation}
and similarly for $\nu_{ii}$.  Implicit in this definition is the assumption that growth occurs in
a way that preserves the relevant attributes of the network (for example, 
the degree distribution).  Suppose that the net in question is the Internet, 
and that $T_{ij}$, representing the average number of links to be randomly followed to connect
two users, is sampled at two different times.
In that case $\nu_{ij}$ reflects the inconvenience in increased transit time associated 
with growth for \emph{old} users, those that were there already at the first sampling, 
whereas $\eta_{ij}$ refers to \emph{all} users, old and new.
We shall see that for some scale-free architectures $\nu_{ij}<\eta_{ij}$. 

Finally, recurrence is a stricter measure of prominence than the mean return time.  Indeed, a finite mean
return time implies recurrence, but diffusion from a node might be recurrent despite having a diverging
mean return time.  We will show that recurrence in scale-free nets may vary from node to node, thus we
shall deem a \emph{node} recurrent (or transient) if diffusion from that particular node is recurrent 
(or transient).
A simple way to decide whether a node is recurrent is to trap all walkers that return to the node and 
consider the survival probability of remaining walkers in the long time asymptotic limit: 
the node is recurrent if the survival
probability tends to zero.

\subsection*{Theoretical techniques and the Complete Graph}

We now demonstrate some basic theoretical techniques for analyzing FPT, as applied to
complete graphs.  In a complete graph of $N$ nodes, K$_N$, all nodes are connected to one another
(there are $\frac{1}{2}N(N-1)$ links).  The complete graphs is 
maximally connected~\cite{remark2}, and serves as a useful yardstick. 
Furthermore, the high symmetry of K$_N$ simplifies its analysis and makes it a perfect start for
pedagogical purposes.

The mean transit time $T_{ij}$ satisfies the equation
\begin{equation}
\label{KTij}
T_{ij}=\frac{1}{N-1} + \frac{N-2}{N-1}(1+T_{ij}).
\end{equation} 
A walker starting off from node $i$ faces $N-1$ equally likely choices (corresponding to the number
of links emanating from $i$).  Only one of the links would lead to node $j$ directly, in one time step.
This choice is represented by the first term on the rhs of Eq.~(\ref{KTij}).  The second term represents
a choice leading to a different node, $h$.  This happens with probability $(N-2)/(N-1)$, which accounts
for the prefactor.  It takes one time step to reach node $h$, and then, the walker is faced with an 
additional mean FPT $T_{hj}$, to reach the target $j$ from node $h$.  However, due to the fact that in
the complete graph all nodes are equivalent, $T_{hj}$ is identical to $T_{ij}$, enabling one to close
the equation.  Solving for $T_{ij}$ we obtain
\begin{equation}
\label{KTijRes}
T_{ij}(N)=N-1.
\end{equation}

The mean return time is analyzed in similar fashion.  It satisfies the simple equation
\begin{equation}
\label{KTii}
T_{ii}=1+T_{ji}.
\end{equation}
A walker starting off from $i$ must step to one of the neighboring sites, $j$, taking one time step
to do so.  This accounts for the first term on the rhs of~(\ref{KTii}).  Once at $j$, the walker needs 
an additional time $T_{ji}$ for returning to $i$ (second term).  However, all sites being equivalent,
$T_{ji}=T_{ij}$.  Hence,
\begin{equation}
\label{KTiiRes}
T_{ii}(N)=N.
\end{equation}
Eqs.~(\ref{KTij}) and (\ref{KTii}) are particular instances of the underlying backward equation
(known also as Dynkin's equation)~\cite{fpp,vk}, a general, powerful approach which is best learnt
by example.  See also Appendix~\ref{matrices}.

For the simple case of the complete graph we can do better than merely deriving the mean FPT;
we can in fact obtain the whole distribution of first passage times.  Let $F_{ij}(t)$ be the probability that
the FPT from $i$ to $j$ is exactly $t$.  It satisfies the equation
\begin{equation}
\label{KP}
F_{ij}(t)=\frac{1}{N-1}\delta_{t,1}+\frac{N-2}{N-1}F_{ij}(t-1),
\end{equation}
where $\delta_{t,1}$ is a Kronecker delta-function.  A walker gets from $i$ to $j$ in one time step with
probability $1/(N-1)$ (first term on rhs), or it gets first to some other site (second term).  
Getting to the other site takes
one time step, so to conclude in $t$ steps the walker needs to complete the trek in $t-1$
steps.

The recursion is best solved by means of the generating function 
\begin{equation}
\label{gf}
{\hat F}_{ij}(x)=\sum_{t=0}^{\infty} F_{ij}(t)x^t.
\end{equation}
Operating on~(\ref{KP}) with $\sum_t x^t$ yields
\begin{equation}
{\hat F}_{ij}(x)=\frac{1}{N-1}x+\frac{N-2}{N-1}x{\hat F}_{ij}(x),
\end{equation}
or
\begin{equation}
{\hat F}_{ij}(x)=\frac{x}{(N-1)-(N-2)x}.
\end{equation}
After Taylor-expanding ${\hat F}_{ij}(x)$, one can read off the probabilities directly:
\begin{equation}
\label{KFij}
F_{ij}(t)=\frac{1}{N-1}\left(\frac{N-2}{N-1}\right)^{t-1}, \qquad t=1,2,\dots
\end{equation}
The distribution of mean transit times is a simple exponential, and it is easily verified that
$T_{ij}=\sum_t tF_{ij}(t)=N-1$, in accord with our previous finding.  It immediately follows that
the distribution of mean return times is exponential as well,
\begin{equation}
\label{KFii}
F_{ii}(t)=\frac{1}{N-1}\left(\frac{N-2}{N-1}\right)^{t-2}, \qquad t=2,3,\dots
\end{equation}
The exponential tail is a generic feature in all finite graphs (regardless of architecture).
For $t\gg T_{ij}$, one expects $F_{ij}(t)\sim\exp(-\alpha t/T_{ij})$ ($\alpha$ is a constant of order unity),
and similarly for $F_{ii}(t)$.  Differences are to be found only in the early time regime, $t\alt T_{ij}$.
Note, however that with $N\to\infty$ this ``early" time regime can be arbitrarily large.

We now turn to the question of recurrence on the complete graph.  To this purpose, imagine an ensemble
of walkers starting off from a node (the ``origin").  
The walkers get trapped whenever they return to the origin, and are
taken out of circulation.  Because all nodes, besides the origin, are equivalent, the probability for finding a
walker in any of the nodes is equal.  Denote the probability for finding a walker in a node (besides the 
origin) at time $t$ by $S_t$.  Clearly, it satisfies
\begin{equation}
S_{t+1}=\frac{N-2}{N-1}S_t,
\end{equation}
for the walker leaves the node with probability one, but walkers from each of the $N-2$ neighboring nodes (all but the origin) step into the site with a probability proportional to $1/(N-1)$, and also proportional to the density of walkers
in the neighboring sites, $S_t$.  The solution is
\begin{equation}
S_t=S_0\left(\frac{N-2}{N-1}\right)^t,
\end{equation}
so we see that in the limit $N\to\infty$ the survival probability is constant: the walker never returns to the
origin, and diffusion on the complete graph is transient.  Note that to test for recurrence
it is necessary to take the limit
$N\to\infty$ \emph{before} considering the long time asymptotic of $t\to\infty$.

Another easy way to see that the complete graph is transient, is from the distribution of FPTs.
We see from Eq.~(\ref{KFii}) that $F_{ii}(t)\to0$ as $N\to\infty$, for any finite $t$.  That is,
as $N\to\infty$ the probability of returning to the origin at any finite time is zero~\cite{remark3}.

With regards to scaling efficiency, the scaling of $T_{ij}$ and $T_{ii}$ with $N$ implies that for
the complete graph
\begin{equation}
\eta_{ij}=1,\qquad\eta_{ii}=1.
\end{equation}
Also, since all sites are equivalent, whether one averages over the preexisting nodes or all nodes
does not make any difference in this case, and
\begin{equation}
\nu_{ij}=1,\qquad\nu_{ii}=1.
\end{equation}

In closing this section, we remind the reader that the complete graph is an extreme, ideal case
of maximal connectivity.  This ideal of connectivity comes at the cost of not being able to maintain
$\av{k}$ constant as $N$ increases.  To do that, most links --- a fraction $1-{\cal O}(1/N)$ --- 
need to be 
removed.  The question is whether one can whittle down the graph in such a way so as to improve
the benchmarks of diffusion considered here.  On the one hand, removing links increases the average
distance between nodes, thus contributing to larger FPTs.  On the other hand, judicious removal
of links can
help a walker focus on following selected paths between the nodes, reducing the risk of getting lost in a plethora of roundabout paths.  It is these two tendencies that must be balanced in designing efficient
nets for transport by diffusion.

\section{Diffusion in other media}
\label{other}

Let us now review how diffusion fares in some  of the better known media, besides
scale-free nets.  Our goal here is to try and understand how diffusion changes from one 
architecture to another, and why.

\subsection*{Regular and fractal lattices}

Regular lattices are one simple way to scale the net in size while maintaining a constant $\av{k}$, 
since the degree of each node is fixed.  Diffusion on regular lattices changes in character with dimensionality.  The mean FPT scales as
\begin{equation}
T_{ii}(N)\sim N,\qquad
T_{ij}(N)\sim\left\{
  \begin{array}{l l}
  N^2,  &d=1,\\
  N,      &d\geq2.
  \end{array} \right.
\end{equation}
In the limit of $N\to\infty$, the distribution of FPTs (whether from a site to itself, or from a site to another
at a finite distance), behaves as
\begin{equation}
F(t)\sim\left\{
  \begin{array}{l l}
 1/t^{3/2},         &d=1,\\
 1/{t\ln^2t},       &d=2,\\
 1/t^{d/2},         &d\geq3.
  \end{array}
\right.
\end{equation}
Note that in $d=1,2$, the mean first passage time $T=\int tF(t)\,dt$ diverges, though diffusion is \emph{recurrent}.
In $d\geq3$, the integral $\int F(t)\,dt<1$ and there is a finite probability that the walker never returns
to the origin: diffusion is \emph{transient} (and the mean FPT diverges as well).

Consider now fractal lattices.  It is possible to construct fractal lattices such that $\av{k}$ is bounded.
A well-known example is the Sierpinski gasket,  
where all nodes other than the three corners
have degree $k=4$.  Diffusion in fractals is anomalous and scales as~\cite{dba}
\begin{equation}
\av{r^2}\sim t^{2/\dw},\qquad \dw>2.
\end{equation}
For regular Euclidean lattices the \emph{walk dimension} $\dw=2$ and diffusion is normal (or Fickian).   
When $\dw$ is larger than the fractal dimension of the underlying medium, $\df$, diffusion is recurrent,
otherwise, if $\dw<\df$, diffusion is transient.  
The mean return time in fractals scales as $T_{ii}(N)\sim N$~\cite{remark4}.  The mean transit time scales as
\begin{equation}
\label{fractal}
T_{ij}\sim\left\{
  \begin{array}{l l}
  N^{\dw/\df}, &\dw>\df,\\
  N,                 &\dw<\df.
  \end{array}
\right.
\end{equation}
For example, for the Sierpinski gasket $\df=\ln3/\ln2$, $\dw=\ln5/\ln2$.  As implied by $\dw>\df$,
diffusion is recurrent, and $T_{ij}(N)\sim N^{\ln5/\ln3}$.

The scaling in~(\ref{fractal}) can be understood in the following way.  
When diffusion is recurrent, the typical time to transit
between $i$ and $j$ is $T_{ij}\sim r_{ij}^{\dw}$.    But the typical distance
between two sites scales as $r\sim N^{1/\df}$, and~(\ref{fractal}) follows.
If diffusion is not recurrent, the target $j$ will
likely not be reached even after the walker's displacement equals $r_{ij}$.  Instead we would typically
wait until each site on the lattice is visited once, on average, before $j$ is reached, and that would take
as many steps as there are nodes in the lattice, or $T\sim N$.  
Regular Euclidean lattices may be subsumed in this description of fractals, with $\df=d$ and $\dw=2$.
Euclidean dimension $d=2$
is a marginal case where $\dw=\df$.  Diffusion turns out to be recurrent in this case.

\subsection*{Erd{\H o}s - R\'enyi graphs}

Another interesting case is that of the Erd\H os - R\'enyi  (ER) random graph~\cite{ER,srb}.  
It consists of a complete
graph K$_N$ where only a fraction $p$ of the links is realized (a fraction $1-p$ of the links is absent).
At $p_c=1/N$
the ER graph undergoes a percolation phase transition: as long as  $p>p_c$ there exists a \emph{giant 
component} of  $N_g$ nodes that forms a finite fraction of the graph, $N_g/N>0$ as $N\to\infty$.  As $p$ approaches $p_c^+$ the giant component disappears, splintering into a multitude of smaller components.

To maintain $\av{k}$ constant, as $N\to\infty$, we must have $p=\mu/N$, $\eta\geq1$. 
(Since there are $N(N-1)/2$ links in K$_N$, this guarantees $\av{k}=\mu$.)  Essentially, we
are looking at the giant component near the percolation threshold.  In this case we know
that $T_{ii}(N_g)\sim N_g$, and
\begin{equation}
T_{ij}(N_g)\sim N_g^{3/2}.
\end{equation}
This result is too related to fractals, as summarized above.  It is well known that percolation in
ER graphs is analogous to percolation in $d=6$ dimensions.  In the latter case, the incipient infinite
percolation cluster (the analogue of the giant component) has fractal dimension $\df=4$, and diffusion
on it is anomalous, with $\dw=6$.  The scaling of $N_g^{3/2}$ found in ER graphs is thus consistent
with $N^{\dw/\df}$ of percolation in $d=6$ dimensions.

\subsection*{Regular trees: the Cayley Tree}

As a last case, before we turn to scale-free graphs, we consider the Cayley tree.
The Cayley tree is a regular tree graph, where each node has fixed degree, $k$.
While normally one considers the  infinite tree, for our purpose we may imagine the
tree to be finite, consisting of $N$ nodes.  Diffusion on the Cayley tree can be analyzed
straightforwardly, by using the underlying backward equation.  One finds that both
$T_{ii}$ and $T_{ij}$ scale linearly with $N$ (for both old or new nodes).  When $N\to\infty$,
diffusion is transient.  An easy way to see this is the following: consider a site on the 
Cayley tree, that we shall take to be the ``origin".  The $k$ neighboring sites constitute shell 1,
and their neighbors (other than the ones on shell 1) are shell 2, etc.  A site on shell $l$
has $k-1$ neighbors on shell $l+1$, but only one neighbor on shell $l-1$.  Thus, diffusion
from shell to shell is biased: the walker steps from $l$ to $l+1$ with probability $(k-1)/k$, as
opposed to probability $1/k$ for stepping back to $l-1$.  This generates a drift away from the origin and 
results in transience.

\subsection*{Summary and outlook}

In all the cases we have seen so far, it would seem that one can have fast FPTs,
or recurrence, but \emph{not} both simultaneously.  Linear scaling of the FPTs with $N$
is the best we have seen, but diffusion seems to be transient (and `finding your way
back home' improbable) whenever that is achieved.  When diffusion is recurrent, it comes
at the expense of a poorer scaling of FPTs with the size of the system: doubling the number of nodes
more than doubles mean transit times.

All the graphs considered above were homogeneous, at least in a statistical sense.  
That means that the distinction
between scaling for old or new nodes (upon growth), mentioned in the introduction, is irrelevant.
Likewise, all nodes in the networks we have seen were of only one type, either transient or recurrent.
We shall now see that in scale-free nets one can achieve both fast FPTs \emph{and} 
recurrence simultaneously, at least for part of the graph.  More precisely, the average FPTs for
\emph{all} nodes in scale-free graphs scales linearly with $N$, but increases \emph{sublinearly} for
the nodes that were already present in the net {before growth}.  
Additionally, the most connected nodes in the
net (the hubs) might be recurrent, though the rest of the net is transient.  
The large inhomogeneity of scale-free graphs (manifest in the degree distribution) 
is partially responsible for these trends.

\section{Diffusion in recursive scale-free nets}
\label{results}

Our discussion of diffusion in scale-free nets is limited to two examples
of deterministic, recursive constructs: a scale-free tree (\tree{}), and a scale-free
graph with loops (\loops{})~\cite{dor_rec}.  They share the same degree distribution, $P(k)\sim k^{-\l}$, 
with exponent $\l=1+\ln3/\ln2$.  A full description of these graphs is given in Appendix~\ref{graphs}.
The two cases symbolize two extremes: \tree{} has no loops (or cycles) --- there is only one possible
path connecting any two nodes;  but \loops{} is full of cycles of all sizes, and the number of alternate routes between two nodes increases as $N^{\ln2/\ln3}$ as the graph grows recursively~\cite{loopy}.  
Random scale-free graphs lie somewhere between these two extremes.

\subsection{FPT for old nodes}

Consider a node $i$ in the $n$-th generation of the scale-free tree, \tree{n}.  
The FPT, for a walker in that node, to exit to \emph{any} of its $k_i$ neighbors, is 1.
Upon growth of the network, to generation $n+1$, the node will double its degree,
from $k_i$ to $2k_i$.  (All the new nodes are of degree 1, see Appendix~\ref{graphs}.)
We will now show that the FPT from $i$ to any of the \emph{old} $k_i$ neighbors,
in the newly grown net \tree{n+1}, is 3.   It follows that the passage time from any node $i \in$ \tree{n}
to any node $j\in$ \tree{n} triples, on average, upon growth to generation $n+1$:
\begin{equation}
\label{Tp_ij}
\Tp_{ij}(N_{n+1})=3T_{ij}(N_n), 
\end{equation}
Since the size of the tree triples asymptotically (as $n\to\infty$), $N_{n+1}=3N_n-3$, this means that 
the scaling efficiency exponent for old nodes is $\nu_{ij}=1$.

\begin{figure}[ht]
  \vspace*{0.cm}\includegraphics*[width=0.25\textwidth]{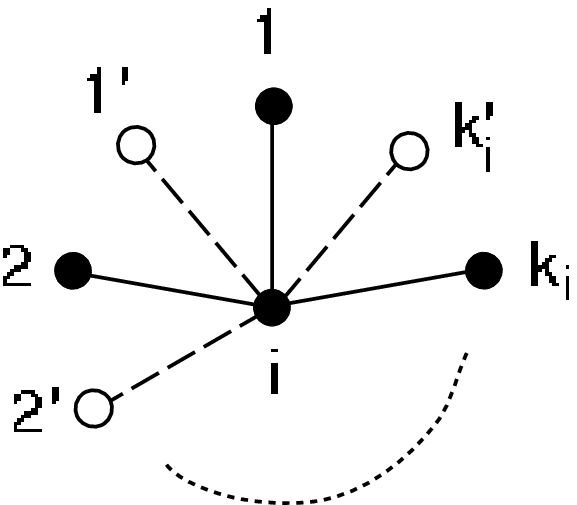}
\caption{Growth in FPTs in going from \tree{n} to \tree{n+1}.  Node $i\in\,$\tree{n} has
$k_i$ neighbors in generation $n$ ($\bullet$), and $k_i$ new neighbors in generation $n+1$ ($\circ$).
The new nodes are of degree one, 
and are connected only to $i$ (broken lines; refer to Appendix~\ref{graphs}).}
\label{TijTree}
\end{figure}

To see the basic scaling, we refer to Fig.~\ref{TijTree}.  Let the FPT for going from node $i$
to any of the old $k_i$ neighbors be $T$.  Let the FPT for going from any of the new $k_i$ neighbors
to one of the old $k_i$ neighbors be $A$.  Then, the underlying backward equations are
\begin{equation}
\begin{split}
&T=\frac{1}{2}+\frac{1}{2}(1+A),\\
&A=1+T,
\end{split}\nonumber
\end{equation}
with solution $T=3$, as stated.   

Consider now the return FPT to node $i$, $T_{ii}$.  Let $\Tp_{ii}$ be the FPT for returning to node $i$ in
\tree{n+1}.  Let $\Tp_{ji}$ be the FPT from $j$ --- an old neighbor of $i\in$ \tree{n} --- to $i$, in \tree{n+1}.
Similarly, let $T_{ii}$ be the FPT for returning to $i$ in \tree{n}, and $T_{ji}$ the FPT from the
same neighbor $j$, to $i$, in \tree{n}.  For \tree{n}, we have the underlying backward equation
\begin{equation}
\label{TiiTree}
T_{ii}=\frac{1}{k_i}\sum_{j=1}^{k_i}(1+T_{ji}), \nonumber
\end{equation}
while for \tree{n+1},
\begin{equation}
\Tp_{ii}=\frac{1}{2}(2)+\frac{1}{2k_i}\sum_{j=1}^{k_i}(1+\Tp_{ji}). \nonumber
\end{equation}
The first term on the rhs indicates the possibility that the walker steps from $i$ to one of the
new neighbors and back, which happens with probability $1/2$ (half of the neighbors are new) and
takes two time steps.
We have already shown that $\Tp_{ji}=3T_{ji}$, which enables us to eliminate 
$\sum T_{ji}$, $\sum\Tp_{ji}$, and yields
\begin{equation}
\label{Tp_ii}
\Tp_{ii}(N_{n+1})=\frac{3}{2}T_{ii}(N_n).
\end{equation}
In other words, $\nu_{ii}=1-\ln2/\ln3<1$.

\begin{figure}[ht]
  \vspace*{0.cm}\includegraphics*[width=0.25\textwidth]{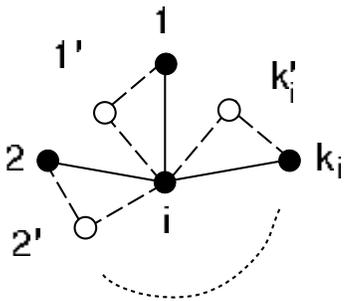}
\caption{Growth in FPTs in going from \loops{n} to \loops{n+1}.  Node $i\in\,\,$\loops{n} has
$k_i$ neighbors in generation $n$ ($\bullet$), and $k_i$ new neighbors in generation $n+1$ ($\circ$).
A new node $j'$ is of degree two, 
and is connected to $i$ and to the old node $j$ (broken lines; refer to Appendix~\ref{graphs}).}
\label{TijLoops}
\end{figure}

\medskip
The scale-free graph with loops, \loops{}, performs even better than that.
In this case the FPT in going from node $i\in$ \loops{n} to node $j\in$ \loops{n}
in the larger graph \loops{n+1} merely doubles:
\begin{equation}
\label{nu_ijLoops}
\Tp_{ij}(N_{n+1})=2T_{ij}(N_n), 
\end{equation}
or $\nu_{ij}=\ln2/\ln3<1$ (compare to $\nu_{ij}=1$, for \tree{}).
To see this basic scaling, we refer to Fig.~\ref{TijLoops} and write the underlying
backward equations for $T$; the FPT for exiting node $i\in$ \loops{n} in the graph \loops{n+1},
and $A$; the FPT for reaching $i$ from any of its new $k_i$ neighbors in generation $n+1$:
\begin{equation}
\begin{split}
&T=\frac{1}{2}+\frac{1}{2}(1+A),\\
&A=\frac{1}{2}+\frac{1}{2}(1+T).
\end{split}  \nonumber
\end{equation}
Eq.~(\ref{nu_ijLoops}) follows from the solution, $T=2$, found upon eliminating $A$.

The equations for $\Tp_{ii}$ are a bit more involved than for the recursive tree \tree{},
for we also need to consider $T_{j'i}$, the FPT for getting from a new neighbor $j'$ back to $i$
(for \tree{}, $T_{j'i}=1$).  We have
\begin{equation}
\begin{split}
&T_{ii}=\frac{1}{k_i}\sum_{j=1}^{k_i}(1+T_{ji}),\\
&\Tp_{ii}=\frac{1}{2k_i}\sum_{j'=1}^{k_i}(1+T_{j'i})+\frac{1}{2k_i}\sum_{j=1}^{k_i}(1+\Tp_{ji}),\\
&T_{j'i}=\frac{1}{2}+\frac{1}{2}(1+\Tp_{ji}).
\end{split} \nonumber
\end{equation}
Again, the relation $\Tp_{ji}=2T_{ji}$ already found above enables one to eliminate all 
the unknown variables, leading to 
\begin{equation}
\label{Tp_ii2}
\Tp_{ii}(N_{n+1})=\frac{3}{2}T_{ii}(N_n), 
\end{equation}
exactly as for \tree{}.  Thus, for returning to the origin we find the same efficiency scaling
exponent $\nu_{ii}=1-\ln2/\ln3$ as for the recursive tree.

We see that in the recursive scale-free nets we have analyzed, the FPT for old nodes
scales more favorably, upon growth, than for all the other cases considered before.  Indeed, 
here $\nu_{ii}<1$ for the first time, and for \loops{} also $\nu_{ij}<1$.
The question arises whether this preferred scaling does not come at the expense of a poorer
scaling for the FPT between new nodes (that constitute the largest part of the net).
We show next that this is not the case, and that $\eta_{ii}=\eta_{ij}=1$. 

\subsection{FPT for all nodes}

Consider  $T_{j'j'}$, the FPT for returning to a new node $j'\in$ \tree{n}, 
a neighbor of $i\in$ \tree{n-1}.  Let $T_1$ be the FPT from $i$ to $j'$, and let $R$ be the FPT
for returning to $i$ (starting from $i$) without ever visiting $j'$.  We have
\begin{equation}
\begin{split}
T_{j'j'}&=1+T_1,\\
T_1&=\frac{1}{k_i}+\frac{k_i-1}{k_i}(R+T_1).
\end{split}  \nonumber
\end{equation}
The second equation here is non-standard and requires elaboration: it is based on
the fact that starting from $i$ the walker would go to $j'$ with probability $1/k_i$ ($k_i$
being the degree of $i$ in \tree{n}), and taking one time step to do so.  With the remaining
probability the walker selects one of the other neighbors, but then takes an average time $R$
to return to $i$, and it still needs time $T_1$ to get to $j'$.

To close these equations we express the FPT for returning to $i$ as
\[
T_{ii}(N_{n})=\frac{1}{k_i}(2)+\frac{k_i-1}{k_i}R.
\]
Eliminating $T_1$ and $R$, this leads to
\begin{equation}
\label{Tjj}
T_{j'j'}(N_{n})=k_iT_{ii}(N_n).
\end{equation}
Finally, using this equation together with the recursion relation of Eq.~(\ref{TiiTree}) and
the fact that $k_i$ doubles in each generation, we find
\begin{equation}
\label{Tj'j'}
T_{j'j'}(N_{n})=2\cdot 3^{n-1}\sim 2N_n.
\end{equation}
Curiously, the result is the same for all new nodes $j'$, despite the fact that they are \emph{not}
equivalent (while they all have degree 1, they are connected to nodes of varying degrees).

We are ready to take averages over \emph{all} nodes.  Iterating~(\ref{TiiTree}), and (\ref{Tjj}), we see that
in \tree{n} there are $2$ nodes with 
$T_{ii}=2\cdot 3^{n-1}/2^{n-1}$, $2$ nodes with $T_{ii}=2\cdot 3^{n-1}/2^{n-2}$, $2\cdot3^m$
with $2\cdot 3^{n-1}/2^{n-2-m}$, \dots, and $2\cdot3^{n-2}$ with $T_{ii}=2\cdot3^{n-1}$.
We thus obtain
\begin{equation}
\av{T_{ii}}_n=\frac{2^{3-n}(9+2^n\cdot3^n)}{5(3+3^{2-n})}\to\frac{8}{15}3^n\sim \frac{8}{5}N_n,
\end{equation}
which shows that $\eta_{ii}=1$.  Basically, the poorer scaling of the new nodes dominates
the average.

Next, consider $T_{ij}$ in \tree{n}, from \emph{any} node $i$ to \emph{any} other node $j$.
New nodes are always of degree $k=1$, and are connected to one of the old nodes (and nothing
else).  Hence, the FPT from a new node $i'$ (a neighbor of the old node $i$) to node $j$ is 
only one time step longer than $T_{ij}$, and does not contribute to the scaling as $N\to\infty$.
We need only worry about $T_{ij'}$, from old node $i$ to new node $j'$ (a neighbor of old node $j$).
The underlying backward equation for this case is
\[
T_{ij'}=T_{ij}+T_{jj'}.
\]
On the other hand,
\[
T_{j'j'}=1+T_{jj'},
\]
which in conjunction with~(\ref{Tj'j'}) yields,
\begin{equation}
T_{ij'}=T_{ij}+2\cdot3^{n-1}-1\sim N_n.
\end{equation}
The last scaling relation follows from~(\ref{Tp_ij}) and $N_n\sim 3^n$.  In summary,
\begin{equation}
\av{T_{ij}}_n\sim N_n,
\end{equation}
and also $\eta_{ij}=1$.

\begin{figure}[ht]
  \vspace*{0.cm}\includegraphics*[width=0.40\textwidth]{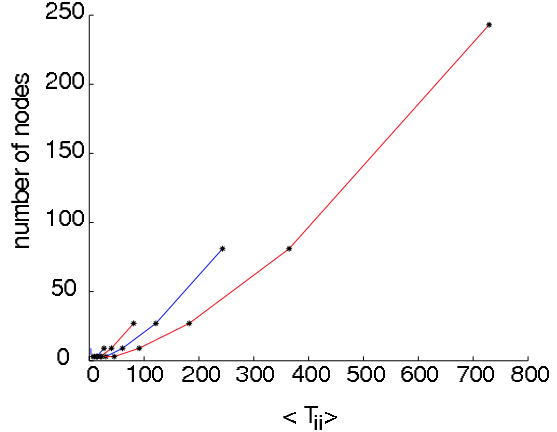}
\caption{Distribution of mean return times in \loops{n}.  Shown are the number of nodes
having a specific mean return time $T_{ii}$, for generations $n=1,2,\dots,6$.
The endpoints of the curves represent new nodes, not present in the previous generation, and follow
the scaling of Eq.~(\ref{Tj'j'loops}).  The remainder of the data points scale in the more advantageous 
way valid for old nodes, Eq.~(\ref{Tp_ii2}).  The scaling of mean return times for \tree{n} is virtually the same.}
\label{FirstLoopTime}
\end{figure}

\medskip
For \loops{} we cannot perform the analogous calculations analytically, namely because of the
way new nodes are connected to \emph{two} old nodes, rather than one, as in \tree{}, barring
us from obtaining $T_{ij'}$ in an obvious way.  Instead, we have written computer code for 
generating and solving the underlying backward equations for all the nodes in \loops{n} 
(see Appendix~\ref{matrices}).  
Memory and time limitations restrict this approach to modest values of $n$, but it becomes quickly
apparent that the scaling is similar to that of \tree{}.  For $T_{ii}$, for example, we have already seen 
that the scaling for old nodes is the same as for \tree{}.  Our numerical results confirm this,
and also reveal that new
nodes scale in the same way as~(\ref{Tj'j'}):
\begin{equation}
\label{Tj'j'loops}
T_{j'j'}(N_n)=3^n\sim 2N_n.
\end{equation}
The end result is that the FPTs for returning to a node in \loops{n} are
$3/2$ as large as the FPTs in \tree{n}, and have the very same distribution (Fig.~\ref{FirstLoopTime}). 
In summary, also for \loops{}, $\eta_{ii}=\eta_{ij}=1$.

\subsection{Recurrence}
The most connected nodes, both in \tree{} and \loops{}, are recurrent.
We can see this  from the following heuristic argument.   Consider the
thermodynamic limit of $N\to\infty$ (or $n\to\infty$).  Consider a node
$i$ in \tree{n} or \loops{n}, that has been added in in the $m$-th generation:
in other words, the network has been iterated $n-m$ times since then.
In each of these iterations, the FPT for returning to $i$ follows the scaling
for old nodes, Eqs.~(\ref{Tp_ii}) and (\ref{Tp_ii2}), so $T_{ii}\sim (3/2)^{n-m}$.
At the same time, the FPT from $i$ to other nodes scales at least as $T_{ij}\sim 3^{n-m}$,
for \tree{n}, or as $T_{ij}\sim 2^{n-m}$, for \loops{n} (the lower bound is realized
for nodes $j$ that were already present in generation $m$ as well).  In either case $T_{ii}/T_{ij}\to0$,
as $n\to\infty$,  suggesting that node $i$ is recurrent.  

In an infinite net only a fraction zero of the nodes is recurrent.  We can relax the definition,
however, for finite $n$ 
and deem node $i$ ``recurrent" if $T_{ii}\ll T_{ij}$,  
or $(3/2)^{n-m}\ll b^{n-m}$,
($b=3$ for \tree{}, and 2 for \loops{}).  The fraction of ``recurrent" nodes is $r=3^m/3^n\equiv(1/3)^{x}$.
Demanding $T_{ii}/T_{ij}=(3/2b)^x\leq1/10$, we get 
$r=2.6\%$ and $0.015\%$ for \tree{} and \loops{}, respectively.

\medskip
For \tree{}, we can show directly that the most connected nodes are recurrent.
The most connected node in \tree{n} has $2^{n-1}$ neighbors, of which $1/2$ are
of degree 1, $(1/2)^2$ are of degree 2, \dots , $(1/2)^m$ are of degree $2^{m-1}$, 
\dots , 1 is of degree $2^{n-2}$, and 1 is of degree $2^{n-1}$.  
Each of these neighbors lie at the root of disconnected branches (they are connected to
one another only through the most connected node).  Using the backward equations,
one can show rather easily that the mean return time to the most connected node,
for a walker that steps into a neighbor of degree $2^m$, is $3^m+1$.  Imagine that
there is a trap, or sink at the origin (the most connected node).  Then, half of the walkers ---
the ones that started off to neighbors of degree 1 --- get trapped after $3^0+1$ time steps;
$(1/2)^2$ get trapped after $3^1+1$ time steps, on average; etc.  After time $t=3^m+1$,
only $(1/2)^m$ of the walkers survive, or
\begin{equation}
S(t)\sim\exp[-({\ln2}/{\ln3})\,t]\to0, \quad \text{as }t\to0,
\end{equation}
proving that the node is recurrent.
A similar argument holds for any node of degree $2^{n-m}$, $n-m<\infty$, $n\to\infty$.

\medskip
Before closing this section, we analyze the FPT distribution for transit between two
hubs A and B in \loops{n}.  Let $F_n(t)$ be the probability that the FPT from hub A to hub B in 
\loops{n} is $t$.  Let $C(t)$ be the probability that the FPT from hub C to hub B in
\loops{n} is $t$.  These satisfy the backward equations:
\begin{equation}
\begin{split}
F_n(t)&=\frac{1}{2}F_{n-1}(t)+\frac{1}{2}\sum_{t'=0}^{t}F_{n-1}(t-t')C(t'),\\
C(t)&=\frac{1}{2}F_{n-1}(t)+\frac{1}{2}\sum_{t'=0}^{t}F_{n-1}(t-t')F_n(t').\\
\end{split} \nonumber
\end{equation}
It is easier to work with the corresponding generating functions, defined as in
Eq.~(\ref{gf}), upon which the convolution terms result in simple products:
\begin{equation}
\begin{split}
{\hat F}_n(x)&=\frac{1}{2}{\hat F}_{n-1}(t)+\frac{1}{2}{\hat F}_{n-1}(x){\hat C}(x),\\
{\hat C}(x)&=\frac{1}{2}{\hat F}_{n-1}(x)+\frac{1}{2}{\hat F}_{n-1}(x){\hat F}_n(x).\\
\end{split} \nonumber
\end{equation}
Considering generation $n=1$ separately, we get the boundary condition $F_0(t)=\delta_{t.1}$,
or ${\hat F}_0(x)=x$, leading to ${\hat F}_n(x)=x/[2^n-(2^n-1)x]$, and
\begin{equation}
F_n(t)=\frac{1}{2^n}\left(\frac{2^n-1}{2^n}\right)^{t-1},\qquad t=1,2,\dots
\end{equation}
Surprisingly, this is the same functional form as Eq.~(\ref{KFij}) for the FPT distribution between nodes
in the complete graph K$_N$, not only in the long time asymptotic limit, which is
expected, but also through the early time regime.  Note, however that the size of the net, $N$,
is here replaced by an effectively smaller size, $2^n\sim N^{\ln2/\ln3}$, emphasizing
the advantages of scale-free nets over other architectures.

\section{Discussion}
\label{discussion}

We have studied diffusion in large networks, under several different architectures, and
have identified clear advantages of scale-free networks over other possibilities.
In one extreme, the complete graph, K$_N$, provides for the shortest possible path between
any two nodes, but offers many alternative paths between the nodes as well.  As a result, the average
transit time between nodes, $T_{ij}(N)$, increases only linearly with $N$.  The ``prominence"
of a node (the ease with which it is re-encountered by a random walker leaving it) is mixed:
the FPT for return to a node scales linearly, $T_{ii}(N)\sim N$, but diffusion is transient (there is a finite
probability to never find the node again).
In the opposite extreme, a one-dimensional chain offers only one path
between any two nodes, but now the distance is large (order $N$).  This results in long
transit times, $T_{ij}(N)\sim N^2$, but increases the prominence of nodes: $T_{ii}(N)\sim N$
(as in K$_N$), and in addition diffusion is now recurrent (a random walker departing
a node will almost surely revisit it).

Other architectures reviewed in this article, besides scale-free nets, lie between these two extremes.
In lattices and fractals, where the distance between nodes is $\sim N^{1/\df}$ ($\df=d$ for 
Euclidean lattices), the situation
is similar to the complete graph when diffusion is transient ($\dw<\df$), and similar to one dimension
when diffusion is recurrent ($\dw>\df$).  For Euclidean {lattices} dimension $d=2$ is a borderline
case separating the two scenarios.  A smaller distance between nodes (as dimensionality
increases) has the further advantage 
of making any \emph{directed} mode of transport, such as diffusion with drift, faster.

Regular trees, such as the Bethe lattice or Cayley tree, boast even smaller distances between nodes
than Euclidean lattices or fractals, of only order $\ln N$, while both $T_{ij}$ and
$T_{ii}$ scale linearly with $N$ (as good as in K$_N$), but diffusion is still transient.  Overall,
diffusion is very similar to that in the complete graph, and the modest logarithmic increase in the distance
between nodes seems a small price to pay for keeping the average degree $\av{k}$ constant 
upon network growth.

The recursive scale-free lattices analyzed in this communication outperform all of the above architectures.
The average FPT for \emph{all} nodes is as good as in K$_N$; $T_{ij}\sim N$, but in
addition, at least for \loops{}, there is an advantage for the nodes \emph{already present} in the network 
before growth; $\Tp_{ij}\sim N^{\nu_{ij}}$, $\nu_{ij}<1$.   (For \tree{} we find $\nu_{ij}=1$, but we suspect
this is the exception rather than the rule, and is probably related to the complete absence of loops.)
The prominence of nodes is, again, at least as good as in K$_N$; $T_{ii}\sim N$, but is advantageous
for old nodes before growth;  $\Tp_{ii}\sim N^{\nu_{ii}}$, $\nu_{ii}<1$ (both for \loops{} and \tree{}).
Also, diffusion from the most connected nodes is recurrent, despite their infinite degree and the essentially
infinite dimensionality of the network.  Finally, the average distance between nodes in scale-free nets is
at least as small as in the Bethe lattice (order $\ln N$) or better, as for random
scale-free nets with degree exponent $2<\lambda<3$ (order $\ln\ln N$).

\medskip
We have examined scale-free nets and showed that they hold some advantages for transport through
diffusion,
but perhaps other architectures are better still.    Also, our study was limited to just two
cases of scale-free recursive nets: many other recursive models exist, but more importantly, 
how do \emph{random} scale-free nets compare to our findings?  These are some of the main open questions
we will be looking at next.

\acknowledgments

We thank S.\ Redner for numerous useful discussions, and NSF grants 
DMS0404778 (EMB) 
and PHY0140094 (DbA) for partial support of this work.

\appendix
\section{Recursive scale-free nets}
\label{graphs}

\begin{figure}[ht]
  \vspace*{0.cm}\includegraphics*[width=0.40\textwidth]{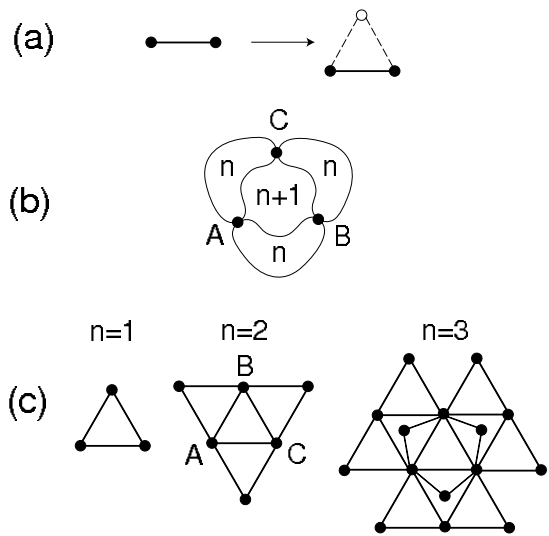}
\caption{The recursive scale-free graph \loops{n}.
(a)~First method of construction: to each link in generation $n$ (solid line) add a new node ($\circ$) of
degree 2, connected to the endpoints of the old link ($\bullet$).  
(b)~Second method of construction: to obtain generation $n+1$, join three copies of generation $n$ at
the hubs (the nodes of highest degree), denoted by A, B, and C in the figure.
(c)~\loops{n} is shown for generations $n=1,2,3$.}
\label{loops}
\end{figure}

\loops{n} is constructed recursively in the following way~\cite{dor_rec}.  The ``seed" in generation $n=1$,
\loops{1}, is identical to K$_3$ (a triangle).  Given generation $n$, one can obtain \loops{n+1}
by adding to each existing link a new node of degree 2, connected to the endpoint vertices of
the link.  Alternatively, \loops{n+1} may be obtained by joining three copies of \loops{n} at the hubs
(the most connected vertices).  The two methods and the resulting graphs are illustrated 
in Fig.~\ref{loops}.  The two different ways for generating \loops{} may be exploited to study
different properties of the graph, as dictated by convenience.   For example, the fact that all
new bonds are of degree $k=2$ (method 1) tells us immediately that $\av{k}=4$, while from the
adjoining of three copies (method 2) we obtain the recursion relation 
\[
N_{n+1}=3N_n-3,
\]
for the order of \loops{n} (the number of nodes in the graph of generation $n$).  This recursion,
coupled with $N_1=3$, yields $N_n=(3^n+3)/2$.  Method 1 also tells us that in going from
generation $n$ to $n+1$ the degree of each existing node doubles,  
while method 2 says that the number of nodes of a given
degree triples (except the hubs).  The degree distribution follows immediately from these two 
observations:  In \loops{n} there are $3^{n-1}$ nodes of degree $2$, $3^{n-2}$ of degree $2^2$,
$3^{n-3}$ of degree $2^3$, \dots, $3^2$ of degree $2^{n-2}$, 3 of degree $2^{n-1}$ and 3 of $2^{n}$.
It follows that $P(k)\sim k^{-\lambda}$, $\lambda=1+\ln3/\ln2$.

\begin{figure}[ht]
  \vspace*{0.cm}\includegraphics*[width=0.40\textwidth]{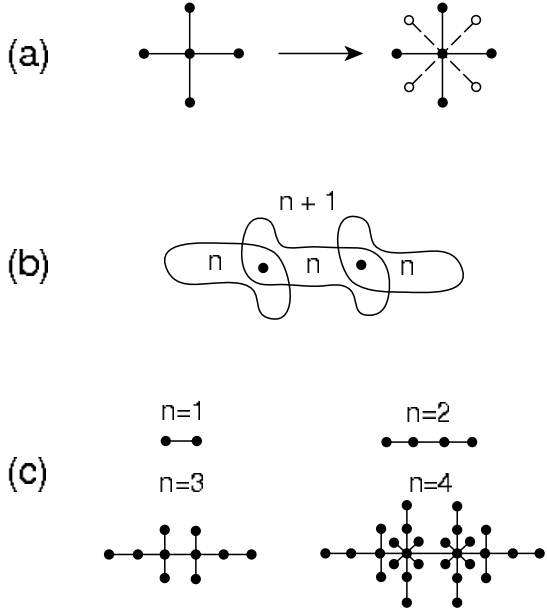}
\caption{The recursive scale-free graph \tree{n}.
(a)~First method of construction: to each node of degree $k$ in generation $n$ (solid line and circles) 
add $k$ new
nodes of degree one, connected only to the old node (broken lines and open circles).  
(b)~Second method of construction: to obtain generation $n+1$, join three copies of generation $n$ at
the hubs (the nodes of highest degree), denoted by A and B in the figure.
(c)~\tree{n} is shown for generations $n=1,2,3,4$.}
\label{trees}
\end{figure}

\medskip
\tree{} is constructed in a similar way.  The seed, \tree{1}, is identical to K$_2$ (a single link connecting
two nodes).  To pass from generation $n$ to $n+1$ one either doubles the degree of each
existing node, by connecting to the node new nodes of degree 1 (method 1), or instead one joins
three copies of generation $n$ at the hubs (method 2).  The two methods of construction and
the first few generations are shown in Fig.~\ref{trees}.  \tree{n} is very similar to \loops{n}:  it has
$2\cdot3^{n-2}$ nodes of degree $1$, $2\cdot3^{n-2}$ of degree $2$,
$2\cdot3^{n-3}$ of degree $2^2$, \dots, $2\cdot3$ of degree $2^{n-3}$, 2 of degree $2^{n-2}$ and 2 of $2^{n-1}$.  The average degree is $\av{k}=2$, and the degree distribution is scale-free, essentially 
the same as that of \loops{}; $P(k)\sim k^{-\lambda}$, $\lambda=1+\ln3/\ln2$.  The most significant
difference between \loops{} and \tree{} is that the latter is a tree --- it has no loops and there is only one
path connecting any two nodes.

\section{Dynkin's Equations for Adjacency Matrices}
\label{matrices}

In this appendix, we present a general form of Dynkin's equations suitable for an arbitrary adjacency matrix.  We take the adjacency matrix $A$, of a graph $G$ with vertices $V=\{v_i\}_{i=1}^N$ and edges $\{(v_i,v_j)\}$, to be written as usual,
\begin{equation}\label{adj}
A_{i,j}=
\left(\begin{array}{ccc}1 & \mbox{ if  there is an edge } & (v_i,v_j) \\
0 & \mbox{ otherwise} & \mbox{ } 
\end{array}\right).
\end{equation}
The Dynkin's equations describe $(T_i)_j$, the expected first transit time from a vertex $i$ to a vertex $j$, and the $N\times 1$ vector ${\mathbf T}_i$ records first transit times for all targets $j$.    We rewrite the graph  Laplacian
\begin{equation}
L=R_A-A, \mbox{ where }R_A=\mbox{rowsum}(A)I,
\end{equation}
where $I$ is the identity matrix, and let
 \begin{equation}
 \ell=R_A^{-1}L=I-R_A^{-1}A.
 \end{equation}
A point-source at vertex $v_i$ is represented by 
 \begin{equation}
 \Delta_i=\mbox{diag }({\mathbf \lambda}_j);\qquad \lambda_j=\delta_{ji},
 \end{equation}
an $N\times N$ matrix of all zeros, except for a $1$ in the $i,i$ position.  It is easy to show that the general form of the  Dynkin's equations may be written as
\begin{equation}
[(I-\Delta)\ell+\Delta]{\mathbf T}=(I-\Delta){\mathbf w},
\end{equation}
where ${\mathbf w}$ is an $N\times 1$ matrix of all 1-entries.
It is expected that this describes a nonsingular system of equations when $G$ has exactly one component, in which case the solution follows,
\begin{equation}\label{Dynkin}
{\mathbf T}=[(I-\Delta)\ell+\Delta]^{-1}(I-\Delta){\mathbf w}.
\end{equation}
Thus, with the graph presented as an adjacency matrix, we can easily compute first transit times, and by solving
Eq.~(\ref{Dynkin}) for each $\delta_i$, for each $i$, we can compute the full matrix of first transit times, $({\mathbf T}_i)_j$.
The expected first return time from a vertex $v_i$ back to itself is easily obtained as
\begin{equation}
T_{ii}=\| {\mathbf T}_i\|_1,
\end{equation}
the column sum of the vector ${\mathbf T}_i$.  In this fashion, from $({\mathbf T}_i)_j$ we can compute the full vector of first return times, $\{T_{ii}\}_{i=1}^N$.
 
 \section{Adjacency Matrices of Self-Similar Graphs}
 \label{adjacency}
 
 In the general random graph case, an adjacency matrix can directly be computed by Eq.~(\ref{adj}).
For the self-similar graphs discussed in this paper, the adjacency matrices can be generated in a neat recursive manner, through the Kronecker-product, $\otimes$.

Consider the sequence of triangular scale-free graphs, \loops{n}, shown in Fig.~\ref{loops}.  Notice that 
\loops{1} has the adjacency matrix,
\begin{equation}
A_1=\left(\begin{array}{ccc}
0 & 1 & 1 \\
1 & 0 & 1 \\
1 & 1 & 0 
\end{array}\right),
\end{equation}
Generally, for recursive networks of the type considered in this paper, there is always a first stage:  we denote the corresponding
adjacency matrix by $A_1$.  Likewise, we denote the 
adjacency matrix of the recursive graph in generation $n$ by  $A_n$.  These $A_n$ can be directly computed by the following algebraic recursion.  Given $A_1$, let
\begin{equation}
A_{n+1}=R_{n+1}\cdot [M \otimes (P_n \cdot A_n \cdot P_n^t)]\cdot R_{n+1}^t.
\end{equation}
The particular form and size of the matrix $M$ dictates how identical copies of $A_n$ will be assembled together.  For example, in the case of \loops{n} we have
\begin{equation}
M=I_3,
\end{equation}
the $3\times3$ identify matrix.  The  $R_n$ are non-square projection matrices.  For example, in the case of \loops{n} the $(n-3)\times n$ matrix $R_n$ is the ``almost diagonal"  matrix written in terms of projections of the $n\times n$ identity matrix $I_n$,
\begin{equation}
R_n=
\left(\begin{array}{c}I_n(1,:)+I_n(n,:) \\
I_n(2:(n/3-1),:)\\
I_n(n/3,:)+I_n((n/3+1),:) \\
I_n((n/3+2):(2n/3-1),:)\\
I_n(2n/3,:)+I_n((2n/3+1),:) \\
I_n((2n/3+2):(n-1),:)
\end{array}\right).
\end{equation}
This pastes the different copies of generation $n$ together, by merging at ``pivot" nodes.  Finally, 
the $P_n$ are $n\times n$ permutation matrices which serve to re-order rows and columns of $A_n$, 
to ensure that the right pivot nodes (the hubs, in the case of
\loops{n}) are picked up for merging next .  

The choice of $R_n$, $M$, and $P_n$, define the construction through the algebra of the Kronecker product for a very large class of recursive graph families.  Furthermore, one could mix-and-match the $R_n$, $M$, and $P_n$ to be different matrices at each stage, to produce an even larger class of graphs. 

In this manner it is straightforward  to produce $A_n$ on a computer for any stage $n$, and together with the general form of the Dynkin's equations discussed in Appendix~\ref{matrices} 
one can easily compute ${\mathbf T}$ and the average FPTs.



\begin{thebibliography}{99}
\bibitem{remark1} One may imagine that adding a new node costs some
energy, money, etc., proportional to the number of links required to connect it,
and that the average rate of expenditure in building the net is constant.
\bibitem{remark2} Consider that in K$_N$ every node is exactly 
one link away from any other node: no other $N$-graph is so tightly connected.
\bibitem{remark3} Survival and FPT are in fact related through $F(t)=-dS(t)/dt$.
\bibitem{remark4} It can actually be shown that $T_{ii}(N)\sim N$
  for a wide class of graphs, under rather general conditions.  See, for example~\cite{perc}.
\end{thebibliography}
\end{document}